# Improving the data quality in the research information systems

Otmane Azeroual[a,b,c]

[a] German Center for Higher Education Research and Science Studies (DZHW), Schützenstraße 6a, Berlin, 10117, Germany

[b] Otto-von-Guericke-University Magdeburg, Department of Computer Science, Institute for Technical and Business Information Systems - Database Research Group, Universitätsplatz 2, 39106 Magdeburg, Germany

[c] University of Applied Sciences HTW Berlin, Department of Computer Science and Engineering, Wilhelminenhofstraße 75A, 12459 Berlin, Germany
Azeroual@dzhw.eu

Mohammad Abuosba[c]

[c] University of Applied Sciences HTW Berlin, Department of Computer Science and Engineering, Wilhelminenhofstraße 75A, 12459 Berlin, Germany
Mohammad.Abuosba@HTW-Berlin.de

*Abstract* — **In order to introduce an integrated research information system, this will provide scientific institutions with the necessary information on research activities and research results in assured quality. Since data collection, duplication, missing values, incorrect formatting, inconsistencies, etc. can arise in the collection of research data in different research information systems, which can have a wide range of negative effects on data quality, the subject of data quality should be treated with better results. This paper examines the data quality problems in research information systems and presents the new techniques that enable organizations to improve their quality of research information.**

## I. Introduction

With the introduction of a research information system, universities and non -university research institutions can provide an up-to-date overview of their research activities, record, process and manage information on their scientific activities, projects and publications, as well as integrate them into their web presence. For scientists, they offer opportunities to collect, categorize and use research information, be it for publication lists, for the preparation of projects, for the reduction of the effort required to produce reports, or for the external presentation of their research and scientific expertise. The data quality is of great importance. Only correct data can provide resilient, useful results and allow for a profound understanding of the research data of research establishment that are always up-to-date. The completeness, correctness, consistency and timeliness of the data are thus essential for successful operational processes. The data errors extend across different areas and weaken the entire research activities of an establishment. Therefore, the aim of this paper is to define and classify the problems of the quality of data that can occur in the research information systems, and then present new techniques that are used to recognize, quantify, and resolve data quality problems in research information systems, to improve their data quality.

## II. Research information system (RIS) - Architecture

A RIS is a central database that can be used to collect, manage and provide information on research activities and research results. The information considered here provides metadata about research activities such as projects, third-party funds, patents, cooperation partners, prices and publications. The RIS architecture usually consists of (see figure 1):

• Data access layer
• Application layer
• Presentation layer

The following figure provides an overview of the individual processes and shows which components belong to which process step:

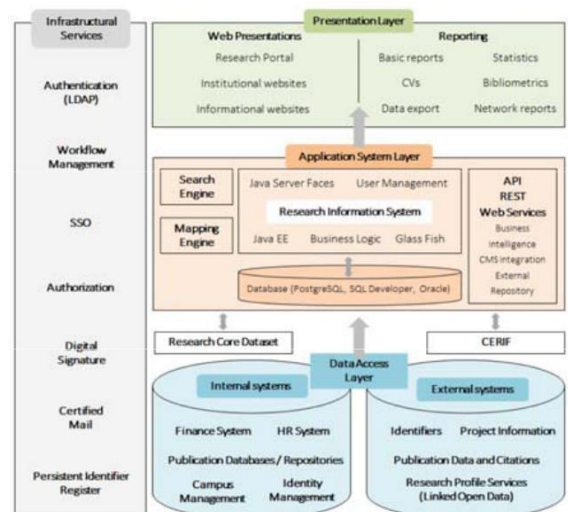

**Figure 1: Architecture of RIS**

The **data access layer** contains the internal and external data sources. This level contains, for example, databases from the administration or publication repositories of libraries, identifiers such as ORCID or bibliographic data from the Web of Science or Scopus etc. The **application layer** contains the research information system and its applications, which merge, manage and analyze the data held at the underlying level.

In the **presentation layer**, the target group-specific preparations and representations of the analysis results are depicted for the user. In addition to various possibilities of reporting, you can also fill portals and websites of the establishment here.

Orthogonal to the described layers, there are the Infrastructural Services, the overlapping services for the entire information systems, such as authentication, authorization, single sign on, etc.

Offers for the standardized collection, provision and exchange of research information in RIS are Research Core Dataset (RCD) and the Common European Research Information Format (CERIF) data model maintained by the non-profit organization euroCRIS version CERIF 2008 1.0 is available. This data model describes the entities as well as their relationship to each other.

### III. Problems of data quality

Collecting data in a database system is a standard process. At each facility, personal data, information about their scientific activities, projects and publications are entered and recorded. The processing and management of this data usually needs to be in a good quality, so that users can get better results.

The quality of data is often defined as the suitability of the data for use in certain required usage objectives, which must be error free, complete, correct, up to date and consistent. Requirements can be set by different stakeholders, in the RIS context e.g. especially by users of a RIS, but also by the RIS administrator. Poor quality data includes data errors (e.g., spelling errors, missing values, incorrect formatting, contradictions and further more). Such quality issues in RIS need to be analyzed and then remedied by data transformation and data cleansing. The following are the typical quality issues of data in the context of a RIS (see figure 2):

● Missing values (features completeness)
● Incorrect information caused by input, measurement
or processing errors (characteristic correctness)
● Duplicates in the dataset (feature redundancy-free)
● Unevenly represented data (feature uniformity)
● Logically contradictory values (consistency characteristic)

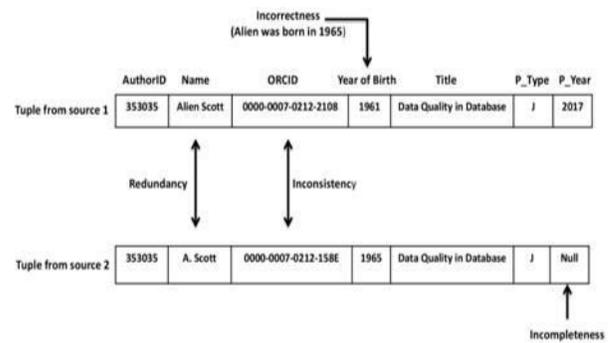

**Figure 2: Examples of data quality problems in RIS**

### IV. Improvement of data quality

Due to the integration of different internal data sources of the establishment and of external sources in research information systems, problems as stated in Chapter 3 have to be overcomes. Now it is important to oppose the causes in this step and to improve the data quality.

The process of identifying and correcting errors and inconsistencies with the aim of increasing the quality of given data sources in RIS, is referred to as data cleansing (or "data cleaning") [8].

Data cleansing includes all necessary activities to clean dirty data (incomplete, incorrect, not up to date, inconsistency, redundant, etc.). The data cleansing process can be roughly structured as follows [3]:

1. Defining and determining the actual problem
2. Find and identify faulty instances
3. Correction of the found errors

Data cleansing uses a variety of specialized methods and technologies within the data cleansing process. [9] subdivides them into the following phases (see figure 3):

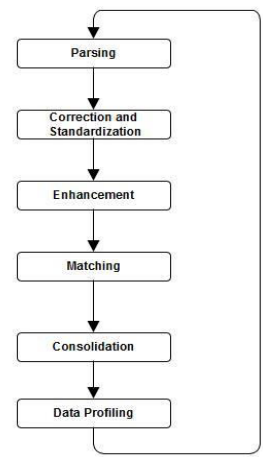

**Figure 3: Data cleansing process**

**Parsing** is the first critical component of data cleansing, helping the user to understand and transform the attributes more accurately. Individual data elements are referenced according to the metadata. This process locates, identifies and isolates individual data elements. For this process, e.g. for names, addresses, zip code and city. The parser Profiler analyzes the attributes and generates a list of tokens from them, and with these tokens the input data can be analyzed to create application-specific rules. The biggest problems here are different field formats that have to be detected.

**Correction & Standardization** is further necessary to check the parsed data for correctness, and then correct it afterwards. Standardization is the prerequisite for successful matching and there is no way around using a second reliable data source. For address data, a postal validation is recommended.

**Enhancement or data enrichment** is the process that expands existing data with data from other sources. Here additional data is added to close existing information gaps. Typical enrichment values are demographic, geographic or address information.

**Matching:** There are different types of matching: for reduplicating, matching to different datasets, consolidating or grouping. The adaptation enables the recognition of the same data. For example, redundancies can be detected and condensed for further information.

**Consolidation (merge):** Matching data items with contexts are recognized by bringing them together.

**Data profiling** is what the data is to be understood, i. E. capture technical structures, analyze them for the purpose of detecting data quality issues, and identify business inconsistencies. Profilers do not describe business rules and do not make any changes. They are only for analyzing the data. Profilers are often used at the beginning of a data analysis, but can also help to better illustrate the results of the analysis.

All of these steps are essential for achieving and maintaining maximum data quality in research information systems. Errors in the integration of multiple data sources in a RIS are eliminated by the clearing up.

The following Table 1 illustrates an example of identifying records with faulty names in a publication list to show how data cleansing processes (clearing up, standardization, enrichment, matching, and merging) can improve the quality of data sources.

The clearing up process adds missing entries, and completed fields are automatically adjusted to a specific format according to set rules.

**Original data before clearing up**

| Data Source | | | | |
|---|---|---|---|---|
| Author ID | Name | ORCID | Birth Date | Address |
| 353035 | Alien Scott | 0000-0007-0212-2108 | 10/25/1965 | 145 F. Concord Street, Orlando, 32801 |
| 353035 | Dr. Alien Scott | 0000-0000-0000-0000 | 25.10.1965 | Concord Street, 32801 145F |
| 353035 | Alien William Scott | 0000-0007-0212-2108 | 652510 | 25 Concord 32801 Street |
| 353036 | A. Scott | | 11/25/56 | 12 Ford Ave 32801 |
| 353036 | Scott Alien | 0000-0007-0212-2108 | 25.11.1965 | |
| | Alien Scoth | 0000000702122108 | 25.10.1956 | Street C., 32801 145F. US |
| 410003 | Olivia Svenson | 0450-1254-3598-F156 | 6-2-1983 | 745-7801 P.B. Las Vegas 29502 |
| 410003 | Svenson Olivia | 045012543598F156 | 1983 | 7801 P.B. Las Vegas 29502 |

**Data after clearing up**

In this example, the missing zip code is determined based on the addresses and added as a separate field. Enrichment rounds off the content by comparing the information against external content, such as demographic and geographic factors, and dynamically expanding and optimizing it with attributes.

| Cleansed Data | | | | | |
|---|---|---|---|---|---|
| Author ID | First | Last | ORCID | Birth Date | Address |
| 353035 | Alien | Scott | 0000-0007-0212-2108 | 1965-10-25 | 32801; FL; Orlando; 145 F. Concord Street |
| 353035 | Alien | Scott | | 1965-10-25 | 32801; FL; Orlando; 145 F. Concord Street |
| 353035 | Alien | Scott | 0000-0007-0212-2108 | | 32801; FL; Orlando; 145 F. Concord Street |
| 353036 | Alien | Scott | | 1965-11-25 | 32801; FL; Orlando; 12 Ford Ave |
| 353036 | Alien | Scott | 0000-0007-0212-2108 | 1956-11-25 | |
| | Alien | Scott | | 1965-10-25 | 32801; FL; Orlando; 145 F. Concord Street |
| 410003 | Olivia | Svenson | 0450-1254-3598-F156 | 1971-02-06 | 29502; NV; Las Vegas; 745-7801 PO Box |
| 410003 | Olivia | Svenson | 0450-1254-3598-F156 | | 29502; NV; Las Vegas; 745-7801 PO Box |

**Data before enrichment**

| Cleansed Data | | | | | |
|---|---|---|---|---|---|
| Author ID | First | Last | ORCID | Birth Date | Address |
| 353035 | Alien | Scott | 0000-0007-0212-2108 | 1965-10-25 | FL; Orlando; 145 F. Concord Street |
| 353035 | Alien | Scott | | 1965-10-25 | FL; Orlando; 145 F. Concord Street |
| 353035 | Alien | Scott | 0000-0007-0212-2108 | | FL; Orlando; 145 F. Concord Street |
| 353036 | Alien | Scott | | 1965-11-25 | FL; Orlando; 12 Ford Ave |
| 353036 | Alien | Scott | 0000-0007-0212-2108 | 1956-11-25 | |
| | Alien | Scott | | 1965-10-25 | FL; Orlando; 145 F. Concord Street |
| 410003 | Olivia | Svenson | 0450-1254-3598-F156 | 1971-02-06 | NV; Las Vegas; 745-7801 PO Box |
| 410003 | Olivia | Svenson | 0450-1254-3598-F156 | | NV; Las Vegas; 745-7801 PO Box |

**Data after enrichment**

The example shows how the reconciliation and merge process runs. Merging and matching promote consistency because related entries within a system or across systems can be automatically recognized and then linked, tuned, or merged.

| Enriched Data | | | | | | |
|---|---|---|---|---|---|---|
| Author ID | First | Last | ORCID | Birth Date | Address | Zip |
| 353035 | Alien | Scott | 0000-0007-0212-2108 | 1965-10-25 | FL; Orlando; 145 F. Concord Street | 32801 |
| 353035 | Alien | Scott | | 1965-10-25 | FL; Orlando; 145 F. Concord Street | 32801 |
| 353035 | Alien | Scott | 0000-0007-0212-2108 | | FL; Orlando; 145 F. Concord Street | 32801 |
| 353036 | Alien | Scott | | 1965-11-25 | FL; Orlando; 12 Ford Ave | 32801 |
| 353036 | Alien | Scott | 0000-0007-0212-2108 | 1956-11-25 | | |
| | Alien | Scott | | 1965-10-25 | FL; Orlando; 145 F. Concord Street | 32801 |
| 410003 | Olivia | Svenson | 0450-1254-3598-F156 | 1971-02-06 | NV; Las Vegas; 745-7801 PO Box | 29502 |
| 410003 | Olivia | Svenson | 0450-1254-3598-F156 | | NV; Las Vegas; 745-7801 PO Box | 29502 |

**Matching**

This example finds related entries for Alien Scott and Olivia Svenson. Despite the similarities between the datasets, not all information is redundant. The adjustment functions evaluate the data in the individual records in detail and determine which ones are redundant and which ones are independent.

| Cleansed Data | | | | | |
|---|---|---|---|---|---|
| Author ID | First | Last | ORCID | Birth Date | Address |
| 353035 | Alien | Scott | 0000-0007-0212-2108 | 1965-10-25 | 32801; FL; Orlando; 145 F. Concord Street |
| 353035 | Alien | Scott | | 1965-10-25 | 32801; FL; Orlando; 145 F. Concord Street |
| 353035 | Alien | Scott | 0000-0007-0212-2108 | | 32801; FL; Orlando; 145 F. Concord Street |
| 353036 | Alien | Scott | | 1965-11-25 | 32801; FL; Orlando; 12 Ford Ave |
| 353036 | Alien | Scott | 0000-0007-0212-2108 | 1956-11-25 | |
| | Alien | Scott | | 1965-10-25 | 32801; FL; Orlando; 145 F. Concord Street |
| 410003 | Olivia | Svenson | 0450-1254-3598-F156 | 1971-02-06 | 29502; NV; Las Vegas; 745-7801 PO Box |
| 410003 | Olivia | Svenson | 0450-1254-3598-F156 | | 29502; NV; Las Vegas; 745-7801 PO Box |

**Consolidation**

The merge makes the reconciled data a comprehensive data set. In this example, the duplicate entries for Alien Scott are merged into a complete record containing all the information.

| Cleansed Data | | | | | |
|---|---|---|---|---|---|
| Author ID | First | Last | ORCID | Birth Date | Address |
| 353035 | Alien | Scott | 0000-0007-0212-2108 | 1965-10-25 | 32801; FL; Orlando; 145 F. Concord Street |
| 353035 | Alien | Scott | | 1965-10-25 | 32801; FL; Orlando; 145 F. Concord Street |
| 353035 | Alien | Scott | 0000-0007-0212-2108 | | 32801; FL; Orlando; 145 F. Concord Street |
| 410003 | Olivia | Svenson | 0450-1254-3598-F156 | 1971-02-06 | 29502; NV; Las Vegas; 745-7801 PO Box |
| 410003 | Olivia | Svenson | 0450-1254-3598-F156 | | 29502; NV; Las Vegas; 745-7801 PO Box |

| Golden Record | | | | | |
|---|---|---|---|---|---|
| Author ID | First | Last | ORCID | Birth Date | Address |
| 353035 | Alien | Scott | 0000-0007-0212-2108 | 1965-10-25 | 32801; FL; Orlando; 145 F. Concord Street |
| 410003 | Olivia | Svenson | 0450-1254-3598-F156 | 1971-02-06 | 29502; NV; Las Vegas; 745-7801 PO Box |

For the frontend of the RIS could be checked with the profiling process, when it comes to the production of statistics,

reports on research data. Profiling makes it easy to assess the overall condition of the data, to identify, prioritize and correct errors, and to remedy the cause of quality issues. Once a profile is created, a facility can respond to quality issues by continually monitoring profile-related parameters.

V. Discussion

The clearing up of data errors in RIS is one of the possible ways to improve existing data quality. Following the defined data cleansing processes, the following developed use cases could be identified in the target system and should serve as a model to show how to detect, quantify, correct and improve them in the case of data errors in research information systems in the establishment.

The following figure introduces the just-mentioned use cases for improving data quality in RIS.

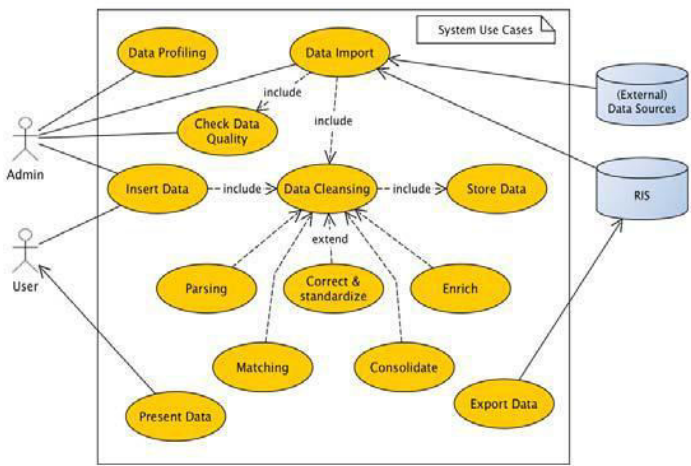

**Figure 4: Use case for improving data quality in the RIS**

The meta process flow can be viewed as shown in the following figure:

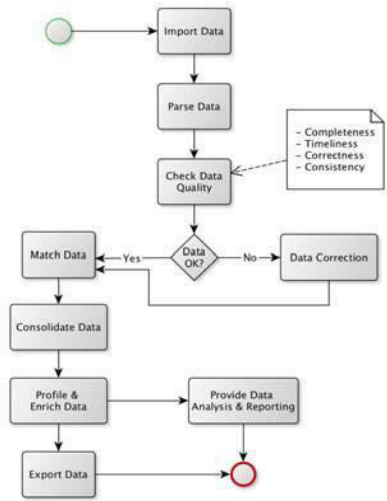

**Figure 5: Main workflow of the process**

The techniques presented by data cleansing help establishments that overcome problems. With these steps every establishment can successfully enforce their data quality.

VI. Conclusion and outlook

In this paper, the question was addressed, which quality problems can occur in research information systems and how to fix and improve them with new techniques or methods, such as data cleansing.

As a result, it were been shown that the improvement of the data quality can be performed at different stages of the data cleansing process in any research information system and that high data quality can be obtained from universities and research institutions to operate e.g. research information systems successfully. In this respect, the review and improvement of data quality are always targeted. The illustrated concept can be used as a basis for the using facilities. It offers a suitable procedure and a use case, on the one hand to be able to better evaluate the data quality in RIS, to be able to prioritize problems better and to prevent them in the future as soon as they occur. On the other hand, these data errors must be corrected and improved with data cleansing. It says: "The sooner quality defects are detected and remedied, the better." Already in the acquisition phase, the author himself or a downstream control authority can correct software errors, such as typing errors, missing values, incorrect formatting, contradictions, etc. To support the universities and all research institutions in the implementation of the data cleansing process, there are numerous tools. With these tools, all facilities can significantly increase the completeness, correctness, timeliness, and consistency of their key data, and they can successfully implement and enforce formal data quality policies. Data cleansing tools are primarily commercial and available for both small application contexts and large data integration application suites. In recent years, a market for data cleansing is developing as a service.

In future work, possible expert interviews and / or quantitative surveys are planned with the universities and research institutions in order to find out how high the data quality is in their research information system and what methods and measures to improve and increase data quality are applied to research information.

## AUTHORS PROFILE

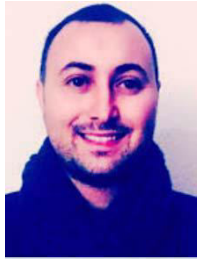

**Otmane Azeroual,** is a researcher at German Center for Higher Education Research and Science Studies (DZHW). After studying Business Information Systems at the University of Applied Sciences Berlin (HTW), he began his Ph.D. in Computer Science at the Institute for Technical and Business Information Systems (ITI), Database Research Group of the Otto-von-Guericke-University Magdeburg and at the University of Applied Sciences (HTW) Berlin in the Department of Computer Science and Engineering. His research interest is in the area of Research Information Systems, Database Systems, Data Quality Management, IT-Security, Cloud Data Management, Data Science, Project Management and Industry 4.0.

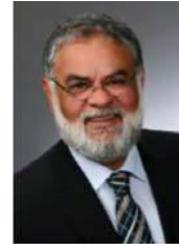

**Prof Dr.-Ing. Mohammad Abuosba,** is a Professor at the Department of Computer Science and Engineering, University of Applied Sciences (HTW) Berlin. His Research areas are Engineering, IT Systems Engineering (focus on Database Systems, Product Data Management), Modeling, Quality Management and Project Management.